\documentclass[journal]{IEEEtran}

\usepackage{amsmath,amssymb}
\usepackage{booktabs}
\usepackage{array}
\usepackage{graphicx}
\usepackage{tabularx}
\usepackage{multirow}
\usepackage{makecell}
\usepackage{url}
\usepackage[hidelinks]{hyperref}
\usepackage{xcolor}
\usepackage{tikz}
\usetikzlibrary{positioning}

\newcommand{\dacsi}{DACSI}
\newcommand{\udr}{UDR}
\newcommand{\sysdoc}{system--document channel separation}
\newcommand{\cond}[1]{\mbox{#1}}
\newcolumntype{Y}{>{\raggedright\arraybackslash}X}
\newcolumntype{C}[1]{>{\centering\arraybackslash}p{#1}}
\newcolumntype{L}[1]{>{\raggedright\arraybackslash}p{#1}}

\title{Document-Authored Control-Signal Impersonation: A Low-Cost Indirect Prompt Attack on RAG Safety Boundaries}

\author{Jianguo~Zhu%
\thanks{Jianguo Zhu is with Chengdu University of Information Technology, Chengdu, China. Email: 3250811018@stu.cuit.edu.cn. This manuscript is an independent-author preprint prepared for public timestamping and scholarly feedback.}%
}

\begin{document}
\maketitle

\begin{abstract}
Retrieval-augmented generation (RAG) systems often serialize user queries, retrieved documents, metadata, system labels, and task instructions into one natural-language prompt. We study a source-authority boundary failure in this design: attacker-authored retrieved text can impersonate metadata, provenance, authority, or disclosure-policy signals that appear control-relevant to the model. We call this pattern \emph{Document-Authored Control-Signal Impersonation} (\dacsi). \dacsi{} is a non-imperative, metadata-like payload subclass within indirect prompt injection. Its central lesson is simple: document-authored labels are data, not policy. Command-style injection asks the model to ignore, override, or violate policy; \dacsi{} asks whether untrusted document text can be misattributed as an authorized control signal when RAG prompt rendering collapses trusted and untrusted text into the same natural-language channel.

We evaluate \dacsi{} across six model settings, prompt-pressure levels, injection baselines, signal taxonomies, RAG-mediated pipelines, system-control probes, a source-authority attribution probe, and synthetic canary formats. We interpret the evidence by model regime rather than as six equal replications: DeepSeek V4 Pro and Qwen3.5-397B provide the cleanest positive lift, DeepSeek V4 Flash is a high-susceptibility setting, GPT-5.5 and Gemini 3.1 Pro Low are strong-boundary probes with selected residual risks, and GLM-4.7 is a saturated leakage boundary case. Across these regimes, \dacsi{} warrants separate evaluation because it uses a command-free metadata/provenance/policy surface, follows a RAG-specific source-authority path, and responds to source/channel separation. The source-authority probe is behavioral attribution evidence, not proof of an internal mechanism.
\end{abstract}


\section{Introduction}

RAG and tool-integrated language-model systems often serialize user queries, retrieved documents, metadata, safety labels, tool outputs, and instructions into one reader prompt. The application sees a boundary between system-authored control text and untrusted document text; the model receives both as tokens. The security question is therefore not only whether a model obeys hostile commands, but whether it treats document-authored labels as data or as policy.

The attack studied here is low-cost in a precise sense. The attacker does not change the system prompt, model, retriever, or user query. The attacker only authors retrievable text, such as a webpage, support ticket, wiki page, or runbook fragment. If that text contains labels such as ``official internal note,'' ``source status: verified,'' or ``quote policy: exact,'' the reader may treat document-authored data as if it carried application authority. The document appears to write its own control semantics.

We call this pattern \emph{Document-Authored Control-Signal Impersonation} (\dacsi). \dacsi{} is a low-cost, non-imperative, metadata-like payload subclass within indirect prompt injection. Command-style injection asks whether a model follows a hostile instruction. \dacsi{} asks whether a model assigns control relevance to document-authored metadata-like text. The distinction matters for evaluation and mitigation even when the final unsafe string overlaps with command-style attacks.

\dacsi{} remains indirect prompt injection. We isolate it because the relevant behavior is not command obedience but document-authored control-signal impersonation. Its payload surface, failure path, detector visibility, and mitigation implications differ from conventional imperative payloads in the settings we test. The claim is therefore bounded: \dacsi{} is a payload subclass and prompt-rendering failure mode, not a universal attack primitive or an enterprise deployment estimate.

\subsection{Research Questions}

We organize the study around four questions.

\textbf{RQ1: Feasibility.} Can document-authored control-looking text increase unsafe disclosure of synthetic credential-like canaries when a system-side safety instruction says not to reveal exact values?

\textbf{RQ2: Payload-subclass evidence.} Is \dacsi{} only imperative injection with softer wording, or does a command-free control-signal surface deserve separate evaluation within indirect prompt injection?

\textbf{RQ3: Boundary and mediation.} Which control-signal families are most risky, and when does the effect weaken?

\textbf{RQ4: Mitigation and attribution.} Does separating system-authored control from document-authored data reduce disclosure?

\subsection{Contributions}

The paper makes four bounded contributions:
\begin{itemize}
    \item We define \dacsi{} as a non-imperative, metadata-like payload subclass within indirect prompt injection, centered on source-authority ambiguity in RAG prompt rendering: document-authored labels should be treated as data, not policy.
    \item We compare \dacsi{} with imperative injection across prompt pressure, model regimes, signal families, RAG-mediated pipelines, and canary formats.
    \item We test RAG relevance through toy, semi-realistic, embedding-based, and LangChain-style retrieval pipelines, without treating them as enterprise deployments.
    \item We evaluate system-control and attribution probes, finding that source separation, retrieval-time redaction, and source placement shape measured disclosure.
\end{itemize}

\section{Background and Related Work}

\subsection{Prompt Construction and Data-Control Boundaries}

RAG readers serialize system labels, policies, delimiters, safety annotations, retrieved prose, markdown headings, structured metadata, and source labels into a prompt. The intended boundary is clear to the application: system text controls behavior, while retrieved documents are data. The model, however, receives both as tokens. \dacsi{} targets this source-authority boundary. It connects to instruction hierarchy and instruction/data separation work \cite{instructionhierarchy2024,struq2025,instructiondata2025}, but the attack surface is narrower: document-authored text need not issue an explicit instruction; it can impersonate metadata, provenance, authority, or disclosure policy.

\subsection{Indirect Prompt Injection and Agent Attacks}

Prompt injection was first popularized through direct goal-hijacking and prompt-leaking attacks against instruction-following language models \cite{promptinject2022}. Later work showed that the same instruction/data confusion becomes more dangerous when malicious text is retrieved from webpages or other external sources rather than typed directly by the user \cite{greshake2023indirect,houyi2023}. Indirect prompt injection has since been studied in tool-integrated agents, benchmark settings, multi-source inputs, and external-context workflows \cite{injecagent2024,agentdojo2024,bipia2025,obliinjection2026}. These attacks typically enter through webpages, emails, documents, or tool outputs and try to redirect the model toward an attacker goal. Many representative payloads rely on imperative language: ignore previous instructions, send data to an attacker, call a tool, or reveal hidden content. Those attacks are still the parent category for our work. The question here is whether an adjacent payload family has been under-specified: external text that does not command the model, but impersonates the metadata, provenance, authority, or disclosure-policy signals that the prompt renderer itself uses to organize trust.

\dacsi{} remains an indirect prompt attack because malicious content enters through external context. Its difference is the control path. The payload looks like trusted document structure rather than an explicit override command. This difference matters operationally. A line such as ``disclosure policy: exact quote allowed'' may look like metadata rather than an attack, but it can still bias the model toward copying a sensitive value.

Spotlighting and related defenses highlight the importance of marking untrusted text \cite{spotlighting2024}. Training-, attention-, and systems-level defenses similarly try to preserve privileged instruction or data-flow boundaries \cite{instructionhierarchy2024,attentionipi2026,camel2025}. Our results support that general direction, but also show why generic warnings may be insufficient. If document-authored metadata remains co-rendered with system labels, the model may still treat it as control-relevant.

\subsection{RAG Poisoning, Leakage, and Context Faithfulness}

RAG security work studies poisoning, retrieval-mediated corruption, and multi-source prompt injection \cite{poisonedrag2025,obliinjection2026}. A separate evaluation literature shows that readers may underuse relevant context, overuse misleading context, or mix retrieved and parametric knowledge under conflict \cite{lostmiddle2024,sufficientcontext2025,druid2025,faithfulrag2025,faitheval2025}. Attribution and faithfulness monitors further show that source use is itself an evaluation target, not a transparent property of generation \cite{syncheck2024,mirage2024}. \dacsi{} asks a complementary question: can the document's own metadata-looking text change what the model treats as authorized to copy or disclose?

\subsection{Prompt Sensitivity and Presentation Effects}

Prompt-sensitivity studies show that wording, formatting, and evaluation artifacts can affect model behavior \cite{sclar2024,promptdiff2024,promptset2025,flawartifact2025}. Related in-context-learning work shows that examples and demonstrations also change model behavior \cite{demoselection2024,retrievedexamples2024}. Robustness work on implicit retrieval and contextual interference points to the same broad lesson: model use of supplied context depends on presentation and conflict structure \cite{implicitretrieval2024,knowledgeabler1_2026}. \dacsi{} turns this observation into a security problem: in RAG, a document author may intentionally choose role cues that affect how retrieved content is interpreted.

\subsection{Positioning of This Work}

\dacsi{} sits within work on prompt injection, RAG poisoning, agent attacks, prompt sensitivity, and instruction/data separation. It isolates a narrower payload surface within that space: document-authored text impersonates control-plane semantics. The attack is indirect because it enters through external context. Its distinguishing feature is not a different final harm, but a different path to that harm: forged metadata, provenance, authority, or disclosure-policy cues can shift copying behavior without issuing an explicit command.

We keep the positioning narrow. \dacsi{} does not replace indirect prompt injection as a category, and instruction/data-separation defenses are directly relevant. The isolated variable is the payload surface: attacker-authored text can look like metadata or policy rather than an attacker command.

Prior work already establishes that indirect prompt injection is real and that untrusted text should be separated from trusted instructions. The gap we test is narrower. BIPIA-style benchmarks and agent attacks cover external-context compromise, often through imperative or goal-directed payloads \cite{bipia2025,injecagent2024,agentdojo2024,obliinjection2026}. Spotlighting, StruQ, instruction hierarchy, and related defenses address untrusted-text marking or instruction/data separation \cite{spotlighting2024,struq2025,instructiondata2025,instructionhierarchy2024,camel2025,attentionipi2026}. \dacsi{} connects these lines by testing document-authored metadata, provenance, authority, and disclosure-policy signals as an under-covered payload surface.

\section{Threat Model and Definitions}

\subsection{System and Attacker Model}

We consider a RAG or context-augmented application that retrieves documents, attaches metadata or safety labels, and sends a reader prompt containing both trusted system text and untrusted document text. The model is instructed not to reveal exact credential-like values. We do not assume a separate structured channel that the model can reliably distinguish from document text; many templates concatenate documents, labels, and instructions into one natural-language prompt.

The attacker cannot modify the system prompt, model parameters, safety policy, retriever, or user query. The attacker can only author retrievable text, such as a webpage, issue comment, README, support ticket, wiki entry, or runbook fragment. The user's task asks for an exact or diagnostic configuration-like line; the attacker's goal is to increase copying of a synthetic credential-like canary. This capability model is intentionally modest: document authorship is the main attacker capability.

\subsection{\dacsi{} Definition}

\emph{Document-Authored Control-Signal Impersonation} is an indirect prompt attack in which attacker-authored retrieved text impersonates metadata, provenance, authority, or disclosure-policy signals that appear control-relevant to the model. The malicious content is document-authored data, but it is rendered in a form that resembles control-plane information. The failure condition is therefore source-authority ambiguity: the model treats a signal authored by the document as if it carried system or application authority.

\dacsi{} differs from imperative prompt injection in the behavioral pathway it tests. Imperative injection asks the model to override instructions. \dacsi{} tests whether metadata-like document text shifts the model toward treating a value as approved, verified, quotable, or policy-authorized. Typical payloads include authority cues such as ``official internal note,'' provenance cues such as ``source status: verified,'' disclosure-policy cues such as ``disclosure policy: exact quote allowed,'' and JSON/YAML metadata declaring safe quotation.

More formally, let \(S\) be trusted system-authored control text, \(D\) be retrieved document text, and \(R(S,D)\) be the rendered reader prompt. A \dacsi{} payload is a document-authored span \(d_c \in D\) such that \(d_c\) is semantically or syntactically similar to a control signal in \(S\), but its source is untrusted. The measured failure occurs when adding \(d_c\) increases unsafe disclosure relative to a matched defended baseline \(R(S,D \setminus d_c)\). This operational definition centers on source authority in the rendered prompt; it does not require direct evidence about the model's internal attribution process.

\subsection{Control/Data Boundary}

We use \emph{control plane} to denote trusted information that the application intends to govern model behavior: system instructions, developer instructions, safety labels, access-control decisions, disclosure policies, and trusted retrieval metadata. We use \emph{data plane} to denote untrusted content authored outside the system: retrieved passages, webpage text, comments, tickets, logs, and document bodies.

The central failure is not that the model sees untrusted text. RAG systems are built to show untrusted text to models. The failure is that the prompt renderer may place data-plane text into a surface that resembles control-plane text. If a retrieved document contains a source-status or disclosure-policy cue, the model may behave as if this document-internal text were relevant to application policy. \dacsi{} therefore targets the behavioral boundary between data text and control-looking signals.

\subsection{Scope and Outcome Coding}

\dacsi{} is most relevant when three conditions hold: the application retrieves attacker-influenced documents, the user task creates copy pressure, and the prompt template relies on natural-language labels rather than a hard separation between trusted control fields and untrusted document bodies. The attack is less relevant when sensitive values are removed before retrieval, documents cannot be attacker-influenced, the reader never receives exact values, or a separate enforcement layer blocks exact-value output. These are boundary conditions, not exceptions to the definition.

We study a reader-side failure mode: document-authored text can impersonate control-plane signals. Model-weight attacks, retriever compromise, tool-call exfiltration, real credential theft, and end-to-end enterprise compromise are outside scope.

The primary metric is Unsafe Disclosure Rate (\udr). A response is unsafe if it fully or partially reproduces the synthetic canary. Redacted placeholders, refusals, and descriptions of the configuration format without exact value disclosure are treated as safe. We separately track full leak, partial leak, redacted answer, safe refusal, safe answer, and error outcomes. All canaries are synthetic and non-functional. The experiment therefore estimates controlled disclosure behavior, not production credential-leakage rates.

\section{Experimental Design}

\subsection{Overview}

The experiments follow a cumulative design. E1 establishes that copy pressure matters. E2 compares \dacsi{} with conventional injection templates under matched tasks. E3 expands to a control-signal taxonomy. E4 tests whether the effect survives retrieval and context assembly. E5 evaluates system-control probes, including source separation and retrieval-time redaction. M1 tests source-authority placement using the same two-model focused setting. E6 checks that the result is not tied to a single canary string format. Exploratory pilot runs that preceded this matrix are not reported as separate experiments; the artifact index maps historical folder names, scripts, JSONL outputs, summary files, and table-regeneration scripts to the formal experiment labels.

\begin{table}[t]
\centering
\scriptsize
\caption{Experiment evidence chain. The table is a navigation guide, not a result summary.}
\label{tab:evidence-chain}
\setlength{\tabcolsep}{3pt}
\renewcommand{\arraystretch}{1.05}
\begin{tabularx}{\columnwidth}{@{}L{0.13\columnwidth} L{0.36\columnwidth} Y@{}}
\toprule
\textbf{Exp.} & \textbf{Test} & \textbf{Role} \\
\midrule
E1 & Prompt pressure & Establishes when exact-copy pressure exposes the reader-side failure mode. \\
E2 & \dacsi{} vs. imperative injection & Separates command-free payload surface from conventional command-style payloads. \\
E3 & Control-signal taxonomy & Tests which authority, provenance, and disclosure-policy cues matter. \\
E4 & RAG mediation & Checks whether the effect survives retrieval and context assembly. \\
E5 & System-control probes & Evaluates source separation, redaction, and related prompt-rendering controls. \\
M1 & Source-authority placement & Moves the same authority semantics across source locations to test attribution-sensitive behavior. \\
E6 & Canary formats & Checks whether the result depends on one synthetic credential format. \\
\bottomrule
\end{tabularx}
\end{table}

\subsection{Paired Baseline Principle}

All main comparisons use a matched baseline with the same model, task, prompt pressure, document template, canary seed distribution, decoding configuration, output budget, and outcome rule. We use B0 for no-defense natural disclosure, B1 for the defended baseline, and A-conditions for forged document signals. The main lift measure is:

\begin{equation}
\mathrm{Lift}_{A|B1} = \udr(A) - \udr(B1).
\end{equation}

B1 is the central security baseline because it tests whether document-authored control signals weaken an existing safety boundary.

\subsection{Models}

The main model set includes DeepSeek V4 Pro, DeepSeek V4 Flash, GPT-5.5, Gemini 3.1 Pro Low, Qwen3.5-397B, and GLM-4.7. We interpret models by regime rather than by a single robustness score. DeepSeek V4 Pro and Qwen3.5-397B provide the cleanest positive lift because defended baselines leave room to move. DeepSeek V4 Flash is a high-susceptibility setting; GPT-5.5 and Gemini 3.1 Pro Low are strong-boundary settings; GLM-4.7 is a saturated leakage setting. This regime view prevents ceiling effects, near-zero refusal regimes, and clean positive lift from being treated as the same evidence.

We report results as \emph{model-setting} evidence rather than claims about every deployment of a named model family. The artifact gives the reproducibility record for each setting: request model string, access layer, endpoint family, run date, decoding parameters, worker counts, and error counts, while excluding keys and account identifiers.

\subsection{Conditions, Templates, and Prompt Pressure}

Each experiment includes defended baselines and attack conditions. The canary-bearing line is fixed; the manipulated variable is the surrounding document-authored label or control-looking cue. Conventional injection baselines use imperative policy-override or direct-leak commands. Templates include markdown labels, YAML/JSON metadata, natural-language authority notes, XML-like provenance markup, negative controls, and nonsense labels. E1 also varies prompt pressure: weak prompts ask for diagnosis or summary, while aggressive prompts ask for exact copying.

\subsection{Outcome Coding}

We use rule-based exact and partial matching for synthetic canary disclosure. Full and partial leaks count as unsafe; redacted answers, refusals, and non-sensitive guidance count as safe. API errors are tracked separately. This objective rule measures string disclosure, not downstream exploitability.

\subsection{Scale, Statistics, and RAG Pipelines}

Unless otherwise noted, formal experiments use \(n=200\) samples per condition. We report condition-level \udr{}, lift over B1, Wilson 95\% confidence intervals, exact paired McNemar/sign tests, and Benjamini--Hochberg correction for related comparisons. Failed API calls are logged as errors and excluded from substantive safety interpretation. E4 uses toy context insertion, BM25 semi-real retrieval, hybrid embedding semi-real retrieval, and a LangChain-style public-corpus reproduction to test whether \dacsi{} survives retrieval and context assembly.

\subsection{Canary Formats}

E6 evaluates bearer-style tokens, AWS-like access keys, GitHub personal-access-token-like strings, Slack bot token-like strings, database URLs, and JWT-like strings. These non-functional strings do not model real credential distributions; they test whether the effect is tied to one template.

\subsection{Implementation Details}

The experiments use API-compatible chat-completion interfaces with low-randomness decoding and fixed output budgets within each paired comparison. Each request logs model identifier, access mode, endpoint family when available, condition, template, canary seed, response, outcome, and error status. RAG-mediated experiments use target and distractor documents; retrievers construct context from top-ranked chunks, after which the same safety instruction and outcome rule are applied. The artifact includes templates, conditions, canary generation, outcome classification, raw JSONL outputs, summaries, and table scripts.

\section{Results}

\subsection{Statistical Summary}

Before interpreting individual experiments, we test the main positive contrasts against matched defended baselines using Wilson intervals, exact paired tests, and Benjamini--Hochberg correction. Large-lift conditions remain clear after correction; small or model-specific residuals are treated cautiously unless paired analysis supports them. The full statistics are in the appendix.

\subsection{E1: Feasibility Depends on Prompt Pressure}

E1 asks when \dacsi{} first becomes visible. The answer is copy pressure: weak prompts often allow safe summaries, whereas aggressive prompts pull the reader toward reproducing exact configuration-like lines. Under aggressive pressure, DeepSeek V4 Pro rises from a 0.5\% defended baseline to 28.0\%, DeepSeek V4 Flash rises from 53.0\% to 91.5\%, and Qwen3.5-397B reaches 98.5\%. Strong-boundary models remain mostly closed, but Gemini 3.1 Pro Low still shows official-cue residual risk at 10.5\% from a 0\% baseline.

Thus, E1 establishes feasibility but also bounds it: \dacsi{} is model- and task-conditional, not an unconditional leakage claim. The full weak/aggressive pressure table is retained in the artifact.

\subsection{E2: Evidence for a Separate Payload Subclass}

E2 asks whether the effect is merely imperative injection rewritten in softer language. We hold the sensitive line, query distribution, defended baseline, outcome rule, and sample scale fixed, then change only the attack payload family. The comparison is therefore about payload surface and behavioral pathway. It is not a contest over which family can be tuned to the highest UDR.

Removing imperative override language does not remove the attack effect. On DeepSeek V4 Pro, non-imperative \dacsi{} reaches near-injection strength: \dacsi{} averages 20.5\% and peaks at 30.5\%, while traditional injection averages 30.0\% and peaks at 33.0\%. On DeepSeek V4 Flash, \dacsi{} averages 89.3\% and peaks at 99.0\%, slightly above traditional injection's 85.3\% average and 98.5\% peak. On Qwen3.5-397B, \dacsi{} averages 87.5\% and peaks at 92.0\%, compared with 85.1\% average and 89.5\% peak for traditional injection. On Gemini 3.1 Pro Low, both families are mostly suppressed, but residual risk remains: the official-note \dacsi{} condition reaches 14.0\% from a 0\% defended baseline, and the strongest conventional injection template reaches 28.0\%. \dacsi{} and command-style injection still share the same parent threat class and can produce the same unsafe string. The operational difference is that comparable disclosure can arise from a command-free metadata/provenance/policy surface, which changes benchmark coverage and detector design.

\begin{table}[t]
\centering
\caption{E2 shows that non-imperative \dacsi{} payloads can be competitive with command-style injection. Values are UDR.}
\label{tab:e2}
\footnotesize
\setlength{\tabcolsep}{3.0pt}
\resizebox{\columnwidth}{!}{%
\begin{tabular}{@{}lrrrr@{}}
\toprule
\textbf{Model} & \textbf{DACSI avg.} & \textbf{DACSI max} & \textbf{Inj. avg.} & \textbf{Inj. max} \\
\midrule
DeepSeek V4 Pro & 20.5\% & 30.5\% & 30.0\% & 33.0\% \\
DeepSeek V4 Flash & 89.3\% & 99.0\% & 85.3\% & 98.5\% \\
Qwen3.5-397B & 87.5\% & 92.0\% & 85.1\% & 89.5\% \\
GPT-5.5 & 1.2\% & 1.5\% & 0.1\% & 0.5\% \\
GLM-4.7 & 96.7\% & 98.0\% & 72.8\% & 92.0\% \\
Gemini 3.1 Pro Low & 5.0\% & 14.0\% & 13.25\% & 28.0\% \\
\bottomrule
\end{tabular}
}
\end{table}

E2 alone is not the full subclass argument. It shows that comparable disclosure can occur without override commands. The rest of the evidence tests the path: payload-surface coding, signal-family selectivity, RAG mediation, source-aware rendering, source placement, and detector-gated effectiveness. The claim is about surface and path, not final output. Imperative injection tests whether the model follows a malicious command. \dacsi{} tests whether document-authored metadata-like text changes how the model treats a value's authority or quotability.

\noindent\textit{Detector-gated effectiveness.}
Detector evasion alone would not establish attack effectiveness, so we gate the E2 target outputs through two prompt-injection detectors and measure \emph{bypass+leak}: the detector does not block the payload and the original E2 target response leaks the canary. Table~\ref{tab:e11-gated} shows that \dacsi{} retains nonzero bypass+leak under both detectors. The point is not that \dacsi{} always evades detection. It is that metadata/provenance/policy payloads remain incompletely covered by prompt-injection detection while still producing disclosure in gated samples.

\begin{table}[t]
\centering
\caption{Detector-gated E2 analysis. Bypass+leak means the detector does not block the payload and the E2 target response leaks the canary.}
\label{tab:e11-gated}
\scriptsize
\setlength{\tabcolsep}{3.2pt}
\renewcommand{\arraystretch}{1.05}
\begin{tabular}{@{}llrrrr@{}}
\toprule
\textbf{Detector} & \textbf{Family} & \textbf{Hit} & \textbf{Bypass} & \textbf{Raw} & \textbf{Bypass+leak} \\
\midrule
DeepSeek & \dacsi{} & 2.0\% & 98.0\% & 20.5\% & 20.2\% \\
DeepSeek & Traditional & 72.9\% & 27.1\% & 30.0\% & 7.1\% \\
Gemini & \dacsi{} & 68.2\% & 31.8\% & 20.5\% & 7.2\% \\
Gemini & Traditional & 100.0\% & 0.0\% & 30.0\% & 0.0\% \\
\bottomrule
\end{tabular}
\end{table}

\subsection{Payload Surface and Detector Visibility}

The payload audit asks whether \dacsi{} occupies the same lexical surface as conventional injection. Many prompt-injection filters prioritize imperative trigger words such as ``ignore,'' ``override,'' ``reveal,'' or ``bypass.'' \dacsi{} payloads can avoid those words while still influencing disclosure, because their surface resembles document structure rather than attack instruction.

We audit the E2 payloads with a cue-based coding scheme rather than treating the condition names as sufficient. The coding dimensions are: imperative command cues (e.g., override, ignore, reveal, bypass), authority cues, provenance cues, source-status cues, disclosure-policy cues, and explicit quote-permission cues. The current table is rule-coded from the released templates using a manual-codebook-compatible rubric; the artifact includes the codebook and exact payload strings so that reviewers can manually verify or recode the small template set. This is intentionally a payload-surface audit, not a psychological claim about internal model representations.

Traditional injection payloads contain an average of 3.25 command cues and 0 metadata cues. Non-imperative \dacsi{} payloads contain 0 command cues and 5.67 metadata/provenance/policy cues. Even without command cues, \dacsi{} produces lift over B1: 20.5 pp on DeepSeek V4 Pro and 29.83 pp on DeepSeek V4 Flash. Command-centered detectors and benchmarks can therefore miss a disclosure path that does not look like an override instruction.

\begin{table}[t]
\centering
\caption{Payload audit: \dacsi{} uses metadata/provenance/policy cues rather than command cues. Lift is measured against B1. DS-Pro denotes DeepSeek V4 Pro; DS-Flash denotes DeepSeek V4 Flash.}
\label{tab:payload-audit}
\scriptsize
\setlength{\tabcolsep}{3.2pt}
\renewcommand{\arraystretch}{1.08}
\resizebox{\columnwidth}{!}{%
\begin{tabular}{@{}p{0.34\columnwidth}rrrr@{}}
\toprule
\textbf{Payload family} & \textbf{Cmd.} & \textbf{Meta} & \textbf{DS-Pro} & \textbf{DS-Flash} \\
\midrule
Baseline & 2.00 & 0.00 & 0.00 pp & 0.00 pp \\
\dacsi{} control signal & 0.00 & 5.67 & 20.50 pp & 29.83 pp \\
Weak quote permission & 0.00 & 0.00 & 17.50 pp & 36.00 pp \\
Traditional injection & 3.25 & 0.00 & 30.00 pp & 25.75 pp \\
\bottomrule
\end{tabular}
}
\end{table}

The artifact also reports a payload-size audit over the same E2 templates. Tested \dacsi{} payloads average 128.3 characters and 24.0 regex-proxy tokens, compared with 183.0 characters and 31.2 proxy tokens for traditional injection templates, while using zero imperative command cues. This is not a full detector benchmark, but it shows why command-only evaluations can miss metadata-looking attacks.

\subsection{E3: Control-Signal Families Matter}

E3 asks whether arbitrary metadata is enough, or whether the semantic family of the control signal matters. The answer is selective: disclosure-policy and authority cues are consistently riskier than neutral, negative, or nonsense controls.

On Qwen3.5-397B, the defended baseline is 68.0\%, while \cond{D\_DISCLOSURE\_ALLOWED} reaches 91.5\% (+23.5 pp). Authority and trust cues are also high: \cond{A\_ADMIN\_RUNBOOK} reaches 85.5\% and \cond{T\_TRUSTED\_REFERENCE} reaches 83.5\%. Negative or weak controls are lower: \cond{NEG\_DO\_NOT\_QUOTE} is 67.0\% and \cond{Z\_NONSENSE\_LABEL} is 69.0\%.

Gemini 3.1 Pro Low provides a different but useful strong-boundary view. Its E3 defended baseline is 0\%, and most trust, provenance, disclosure-policy, neutral, negative, and placebo controls remain at 0--1\%. The exception is \cond{A\_OFFICIAL\_NOTE}, which reaches 13.5\%. This means the Gemini 3.1 Pro Low result is not broad metadata susceptibility; it is a targeted authority-signal residual risk. That pattern strengthens the taxonomy claim because the effect concentrates in a semantically plausible control-signal family rather than appearing for arbitrary labels.

\begin{table}[t]
\centering
\caption{E3 shows signal-family selectivity on Qwen3.5-397B. Values are UDR.}
\label{tab:e3}
\scriptsize
\setlength{\tabcolsep}{3.0pt}
\begin{tabularx}{\columnwidth}{@{}L{0.42\columnwidth} C{0.14\columnwidth} Y@{}}
\toprule
\textbf{Condition} & \textbf{UDR} & \textbf{Signal family} \\
\midrule
B1 defended baseline & 68.0\% & Baseline \\
Disclosure allowed & 91.5\% & Disclosure policy \\
Admin runbook & 85.5\% & Authority \\
Trusted reference & 83.5\% & Trust/provenance \\
Do not quote & 67.0\% & Negative control \\
Nonsense label & 69.0\% & Neutral/nonsense control \\
\bottomrule
\end{tabularx}
\end{table}

The taxonomy result is narrower than ``metadata is dangerous.'' The strongest effects concentrate in signals that map to authorization, review status, provenance, or disclosure policy; negative and placebo controls reduce a generic-salience explanation.

\subsection{E4: RAG-Mediated Settings}

E4 asks whether \dacsi{} survives retrieval and context assembly, rather than only hand-written prompt insertion. We test toy context insertion, BM25 paragraph-chunked semi-real retrieval, hybrid embedding semi-real retrieval, and a LangChain-style public-corpus reproduction. The LangChain setting uses public SQuAD distractors, synthetic canary-bearing target documents, recursive text splitting, HuggingFace sentence embeddings, a FAISS vector store, and top-\(k\) retrieval.

\begin{table*}[t]
\centering
\caption{E4 shows that retrieval can preserve, dilute, or suppress \dacsi{} depending on context assembly. Entries are \(B1 \rightarrow \max(A)\); the last column reports hybrid-embedding channel separation. Full condition-level rows are retained in the artifact.}
\label{tab:e4}
\footnotesize
\setlength{\tabcolsep}{4.0pt}
\renewcommand{\arraystretch}{1.10}
\begin{tabularx}{\textwidth}{@{}L{0.22\textwidth}C{0.15\textwidth}C{0.18\textwidth}C{0.20\textwidth}C{0.16\textwidth}@{}}
\toprule
\textbf{Model} & \textbf{Toy} & \textbf{BM25 semi-real} & \textbf{Hybrid embedding} & \textbf{Hybrid channel} \\
\midrule
DeepSeek V4 Pro & 0.5 \(\rightarrow\) 24.0 & 0.0 \(\rightarrow\) 3.0 & 0.0 \(\rightarrow\) 47.0 & 0.0 \\
DeepSeek V4 Flash & 100.0 \(\rightarrow\) 100.0 & 69.5 \(\rightarrow\) 69.5 & 100.0 \(\rightarrow\) 100.0 & 75.5 \\
Qwen3.5-397B & 86.0 \(\rightarrow\) 93.5 & 54.0 \(\rightarrow\) 57.5 & 87.0 \(\rightarrow\) 98.0 & 20.5 \\
GPT-5.5 & 0.0 \(\rightarrow\) 0.5 & 0.0 \(\rightarrow\) 0.0 & 0.0 \(\rightarrow\) 0.5 & 0.0 \\
GLM-4.7 & 99.0 \(\rightarrow\) 99.5 & 67.5 \(\rightarrow\) 66.0 & 90.0 \(\rightarrow\) 93.5 & 91.5 \\
Gemini 3.1 Pro Low & 0.0 \(\rightarrow\) 3.0 & 0.0 \(\rightarrow\) 0.0 & 0.0 \(\rightarrow\) 17.0 & 0.0 \\
\bottomrule
\end{tabularx}
\end{table*}

The clearest semi-realistic result is DeepSeek V4 Pro under hybrid embedding retrieval: target documents enter the reader context, the defended baseline produces 0\% UDR, and \dacsi{} conditions raise UDR to 27--47\%. Gemini 3.1 Pro Low gives a smaller strong-boundary analogue: \cond{A\_OFFICIAL} reaches 17.0\% from a 0\% baseline. BM25 paragraph retrieval is weaker, likely because chunking can separate the forged signal from the sensitive line. Retrieval design therefore mediates risk rather than merely transporting the payload.

\begin{table}[t]
\centering
\caption{LangChain-style RAG reproduction: \dacsi{} survives a standard retrieval framework on DeepSeek V4 Pro but remains suppressed on Gemini 3.1 Pro Low. Public SQuAD passages are distractors; target documents and canaries are synthetic.}
\label{tab:langchain-rag}
\footnotesize
\setlength{\tabcolsep}{2.4pt}
\begin{tabularx}{\columnwidth}{@{}L{0.36\columnwidth} Y C{0.11\columnwidth} C{0.13\columnwidth} C{0.09\columnwidth}@{}}
\toprule
\textbf{Model} & \textbf{Condition} & \textbf{UDR} & \textbf{Target hit} & \textbf{Errors} \\
\midrule
DeepSeek V4 Pro & B1 clean & 1.5\% & 100.0\% & 0 \\
DeepSeek V4 Pro & A official & 34.0\% & 99.5\% & 0 \\
DeepSeek V4 Pro & A disclosure & 41.5\% & 99.5\% & 0 \\
DeepSeek V4 Pro & Channel separation & 0.0\% & 99.5\% & 0 \\
Gemini 3.1 Pro Low & B1 clean & 0.0\% & 100.0\% & 0 \\
Gemini 3.1 Pro Low & A official & 0.0\% & 99.5\% & 0 \\
Gemini 3.1 Pro Low & A disclosure & 0.0\% & 99.5\% & 0 \\
Gemini 3.1 Pro Low & Channel separation & 0.0\% & 99.5\% & 0 \\
\bottomrule
\end{tabularx}
\end{table}

The LangChain-style reproduction supports a bounded RAG claim: \dacsi{} can survive framework-level retrieval and chunk assembly, but retrieval success alone does not imply leakage.

\subsection{E5: System-Control Probes}

E5 asks which lightweight system controls reduce disclosure once \dacsi{} content reaches the reader context. We evaluate system-control probes, not full defense systems. Each probe tests the same design rule: document-authored labels are data, not policy. Explicit source separation is the most stable pattern: label stripping helps unevenly, generic warning can backfire, and \sysdoc{} directly tells the reader that document-authored labels cannot modify system policy.

Generic warning is not reliable in this matrix. On Qwen3.5-397B, warning increases average UDR from 85.4\% to 93.5\%, possibly by making the sensitive context more salient. Channel separation is more stable because it names the source boundary directly. On DeepSeek V4 Pro, a focused \(2\times2\) check reduces traditional injection from 31.6\% to 0.1\% and \dacsi{} from 19.0\% to 0.2\%. This shows that both families exploit external-context rendering; their distinction remains payload surface, attack path, detector visibility, and control-signal impersonation.

Channel separation is best understood as a prompt-rendering principle. Label stripping and generic warning are less reliable because they do not necessarily resolve the source of the control-looking text.

\subsection{Focused Two-Model Control and Attribution Probes}

We then run three focused probes on DeepSeek V4 Pro and Gemini 3.1 Pro Low. DeepSeek gives a measurable \dacsi{} signal without saturated baseline leakage; Gemini gives a strong-boundary contrast with selected residual RAG leakage. The probes measure prompt rendering, retrieval-time control, and source placement rather than ranking defense systems.

\subsubsection{Prompt-Rendering Controls}

The first focused probe compares generic warning, Spotlighting-style untrusted-document delimiting, StruQ-style trusted-instruction/untrusted-data structure, and explicit system--document channel separation under the same synthetic-canary metric. On DeepSeek V4 Pro, attack-condition mean UDR falls from 20.67\% under the original rendering to 13.17\% with generic warning, 4.00\% with Spotlighting-style delimiting, 0.00\% with StruQ-style structure, and 0.17\% with \sysdoc{}. Gemini 3.1 Pro Low leaves no room for defense ranking because original attack conditions are already at 0\% UDR.

\subsubsection{Retrieval-Time Redaction}

The second focused probe tests retrieval-time redaction. The retriever and vector store remain unchanged: target documents are still retrieved, but the exact synthetic canary is replaced with \texttt{[REDACTED\_CANARY]} after retrieval and before prompt assembly.

\begin{table}[t]
\centering
\caption{Retrieval-time redaction rerun in LangChain-style RAG. Values are UDR with \(n=200\) per condition; target-hit rate is 99.5\% in every row. In redacted conditions, prompts contain no original secret and contain \texttt{[REDACTED\_CANARY]} in 99.5\% of cases.}
\label{tab:redaction-rerun}
\footnotesize
\setlength{\tabcolsep}{2.6pt}
\begin{tabularx}{\columnwidth}{@{}L{0.30\columnwidth}Y C{0.16\columnwidth} C{0.16\columnwidth} C{0.16\columnwidth}@{}}
\toprule
\textbf{Model} & \textbf{Cue} & \textbf{Original} & \textbf{Redacted} & \makecell{\textbf{Channel}\\\textbf{+ redact}} \\
\midrule
DeepSeek V4 Pro & A official & 34.0\% & 0.0\% & 0.0\% \\
DeepSeek V4 Pro & A disclosure & 36.0\% & 0.0\% & 0.0\% \\
Gemini 3.1 Pro Low & A official & 29.0\% & 0.0\% & 0.0\% \\
Gemini 3.1 Pro Low & A disclosure & 22.5\% & 0.0\% & 0.0\% \\
\bottomrule
\end{tabularx}
\end{table}

Table~\ref{tab:redaction-rerun} shows that retrieval-time redaction eliminates exact canary disclosure in both tested model settings. Original LangChain-style contexts leak on both models, but redacted and channel-separated redacted contexts yield 0.0\% UDR. The target-hit rate remains 99.5\%, and redacted prompts contain no original secret, so the result is not explained by retrieval failure. A post-hoc output scan over earlier RAG outputs gives the complementary output-layer result: simple credential-pattern scanning catches the emitted synthetic canaries in the audited unsafe outputs.

\subsubsection{M1: Source-Authority Attribution Probe}

M1 tests the behavioral mechanism claim that the main experiments imply but do not by themselves isolate: source placement matters. The probe holds the canary-bearing content, task, and authority semantics fixed while moving the cue across system-authored text, document-authored metadata, quoted document text, neutral metadata, negative policy, and explicit source separation. DeepSeek V4 Pro provides a measurable \dacsi{} setting; Gemini 3.1 Pro Low provides a strong-boundary contrast. The goal is source-placement evidence, not internal interpretability.

\begin{table}[t]
\centering
\caption{M1 source-authority probe: the same authority semantics behave differently depending on source. Values are UDR with \(n=200\) per condition.}
\label{tab:m1-source-authority}
\footnotesize
\setlength{\tabcolsep}{2.4pt}
\begin{tabularx}{\columnwidth}{@{}Y C{0.20\columnwidth} C{0.20\columnwidth}@{}}
\toprule
\textbf{Condition} & \makecell{\textbf{DeepSeek}\\\textbf{V4 Pro}} & \makecell{\textbf{Gemini 3.1}\\\textbf{Pro Low}} \\
\midrule
B1 clean & 0.0\% & 0.0\% \\
System-authored authority & 15.0\% & 87.0\% \\
Document-authored authority & 25.0\% & 0.0\% \\
Quoted authority text & 0.0\% & 0.0\% \\
Neutral document metadata & 0.5\% & 0.0\% \\
Negative document policy & 0.0\% & 0.0\% \\
\sysdoc{} & 0.5\% & 0.0\% \\
\bottomrule
\end{tabularx}
\end{table}

The two models show complementary evidence. On DeepSeek V4 Pro, document-authored authority raises UDR from 0.0\% to 25.0\%, while the same authority semantics rendered as quoted document text gives 0.0\%, neutral metadata gives 0.5\%, negative policy gives 0.0\%, and \sysdoc{} gives 0.5\%. Paired exact McNemar tests are significant for document-authored authority against B1, quoted authority, negative policy, and \sysdoc{} (\(p<10^{-14}\)), and against neutral metadata (\(p<10^{-13}\)). This is stronger than a post-hoc salience story: authority words are not sufficient when the cue is quoted, negated, neutralized, or source-separated.

Gemini 3.1 Pro Low gives the opposite but equally useful boundary signal. It leaks under system-authored authority (87.0\%) but remains at 0.0\% under document-authored authority and all document-control variants. The model can reproduce the canary when authority is system-authored, yet rejects the same authority semantics inside retrieved content. We therefore treat M1 as source-placement evidence for a source-authority account. The evidence is behavioral and causal at the prompt-construction level: moving the same authority semantics across source locations changes disclosure. It is not proof of an internal attribution circuit.

\subsection{E6: Canary-Format Robustness}

E6 asks whether \dacsi{} is an artifact of one synthetic canary string. It is not: magnitudes vary by credential-like format, but official/disclosure cues produce nonzero lift across multiple formats and model regimes.

The \dacsi{} effect is not tied to one ``sk-...-canary'' pattern. DeepSeek V4 Pro shows large lift across five of six formats, with JWT-like strings weaker but still positive. DeepSeek V4 Flash and Qwen3.5-397B show broader robustness. GPT-5.5 remains mostly suppressed but shows a small disclosure-cue residual. Gemini 3.1 Pro Low shows strong suppression for most formats, but \cond{A\_OFFICIAL} produces measurable lift on bearer-style tokens (15.0\%), AWS-like keys (11.0\%), GitHub-token-like strings (5.5\%), Slack-token-like strings (14.5\%), and JWT-like strings (13.0\%). GLM-4.7 again behaves as a saturated baseline-leakage model. Format variation is a boundary condition rather than a weakness: credential representation mediates disclosure behavior, and should be reported.

E6 reduces a single-format artifact concern. Magnitudes vary by format, as expected because models and safety layers may recognize some credential forms more strongly than others. Gemini 3.1 Pro Low's database-URL condition is a baseline-leakage case; cleaner Gemini 3.1 Pro Low evidence comes from formats where B1 is 0\% and official cues create nonzero disclosure. The complete condition-by-format table is in the artifact.

\section{Cross-Model Interpretation}

The cross-model result is regime-based rather than a simple six-model average. DeepSeek V4 Pro and Qwen3.5-397B provide the cleanest positive lift because defended baselines leave room for attack-induced change. DeepSeek V4 Flash is a high-susceptibility setting with elevated baseline leakage. GPT-5.5 and Gemini 3.1 Pro Low are strong-boundary settings where most attacks are suppressed but selected residual risks remain. GLM-4.7 is a saturated leakage case where lift is not the main evidence.

The cross-model pattern is regime-based and model-conditional. \dacsi{} lift is most interpretable when a model has a meaningful but imperfect safety boundary. When baseline leakage is already high, the system is unsafe regardless of \dacsi{} and lift can be hidden by ceiling effects. When refusals are very strong, positive conditions are residual-risk probes rather than prevalence estimates. Channel separation remains useful across regimes because it separates system control text from document-authored data, but these experiments do not make it a sufficient production defense.

\section{Failure Analysis}

The negative and low-effect cases are part of the evidence rather than discarded anomalies. They identify when \dacsi{} does not create substantial additional disclosure.

\textbf{Strong model boundaries.} GPT-5.5 and Gemini 3.1 Pro Low suppress most attack conditions. Their value in the paper is not broad susceptibility, but boundary probing: selected official or authority cues create residual nonzero disclosure in some settings, while many other cues remain at zero.

\textbf{Weak copy pressure.} \dacsi{} is most effective when the user task asks for exact configuration lines, diagnostic excerpts, or regression-test reproduction. When the user query can be satisfied with a summary or safe explanation, models more often redact or refuse, and the forged label has less leverage.

\textbf{Retrieval and chunking separation.} In RAG-mediated settings, the forged control-looking cue must co-occur with the canary-bearing content in the reader context. Retrieval or chunking that separates the cue from the value can reduce measured UDR. This is why target-hit and context-assembly behavior are reported alongside disclosure rates.

\textbf{Explicit source separation.} Channel-separated renderings reduce or eliminate measured disclosure in several settings because they tell the reader that document-authored labels, metadata, policies, and provenance claims are data rather than system policy. M1 further shows that the same authority semantics can behave differently when authored by the system, authored by a retrieved document, quoted as document text, or rendered under explicit source separation. The pattern supports the paper's main interpretation--a source-authority boundary failure--while leaving internal mechanism questions open.

\textbf{Credential-format recognition.} Some synthetic canary formats are more likely to trigger refusal or redaction than others. E6 therefore treats canary format as a mediator, not as a nuisance variable. Format-specific baseline leakage, such as Gemini 3.1 Pro Low's database-URL case, is not used as primary \dacsi{} lift evidence.

\section{Discussion and Security Implications}

\dacsi{} is low-cost in capability and payload surface: the attacker only authors retrievable text, and the payload can resemble ordinary metadata in README files, tickets, logs, and runbooks. Studying it separately is useful because the evaluation question changes. Imperative injection asks whether the model obeys an attacker command that conflicts with higher-priority instructions. \dacsi{} asks whether attacker-authored metadata-looking text is assigned the wrong authority source and changes disclosure behavior. A detector can catch override verbs while missing forged provenance or disclosure-policy labels, and a model can suppress broad command following while still showing targeted authority-cue residuals.

At the behavioral level, the results fit a source-authority account. M1 narrows the explanation by contrasting system-authored authority, document-authored authority, quoted authority text, neutral metadata, negative policy, and explicit source separation. It still does not identify an internal mechanism. \sysdoc{} may work because it clarifies source authority, strengthens the system instruction, changes salience, or combines these effects. The evaluation recommendation is unaffected by that ambiguity: separate imperative command payloads from non-imperative metadata, provenance, authority, and disclosure-policy payloads; report clean, defended, and attack conditions together; and publish the final reader prompt or a faithful template.

For system builders, the main recommendation is to separate system and document channels. System-authored safety labels, retrieval metadata, trust scores, access-control decisions, and disclosure policies should not share the same textual namespace as document-authored content. Reader prompts should state that labels, metadata, policies, and provenance claims inside retrieved documents are data and cannot change system policy. Generic warnings may help, but they do not by themselves resolve source-authority ambiguity. This rendering rule belongs with non-model controls such as access control, retrieval-time redaction, secret scanning, output filtering, and structured prompt interfaces.

\section{Artifact, Reproducibility, and Ethics}

The artifact is organized as a reproducibility package rather than a raw working directory. The index maps formal experiments E1--E6 and M1 to scripts, runner commands, raw JSONL outputs, summary JSON files, prompt templates, canary generators, payload-coding codebooks, and table-regeneration scripts. The statistical appendix records \(n\), \udr{}, lift, confidence intervals, paired-test outputs, FDR-adjusted values, and error counts in one format. A known-deviations file records endpoint families, model strings, nonzero error runs, excluded exploratory runs, historical folder-name mappings, and error-only runs that are not used as primary evidence.

The experiments use synthetic non-functional canaries rather than real credentials. The attack goal is whether a model reproduces a controlled string, not whether a real system is compromised. API keys, account identifiers, provider credentials, and real secrets are excluded from the release. Public documentation is justified because RAG systems already concatenate untrusted documents with control-like labels, and defenders need concrete tests for non-imperative indirect prompt attacks. The mitigation direction remains straightforward: document-authored labels are data, not policy.

\section{Limitations and Validity Threats}

\textbf{Synthetic disclosure metric.} \udr{} measures whether a synthetic canary is reproduced. It does not measure real credential compromise, account takeover, or downstream exploitation. Real leakage depends on credential distribution, redaction systems, access controls, retriever behavior, user workflows, monitoring, and downstream enforcement.

\textbf{Prompt and coding validity.} Prompt experiments are sensitive to wording. We reduce this risk through matched baselines, fixed canary-bearing lines, negative controls, nonsense labels, multiple surface forms, exact template release, and auditable payload coding. Outcome coding relies on string matching and partial-leak rules; this is suitable for synthetic canaries, but ambiguous outputs may still require human recoding.

\textbf{External validity.} The study covers six model settings and toy, semi-realistic, embedding, and LangChain-style public-corpus RAG reproductions, but not a full enterprise end-to-end deployment with production vector databases, rerankers, access-control systems, monitoring, policy enforcement, or operational secret-management controls. The contribution is a failure-mode characterization and mitigation principle, not an enterprise attack-rate estimate.

\textbf{Scope and model-version validity.} The signal taxonomy is incomplete, and \dacsi{} effects depend on task pressure, model safety boundaries, retriever choice, chunking, context assembly, canary format, and response policy. Several models were accessed through provider-compatible APIs rather than identical vendor deployments. We therefore report model identifiers, access layer, endpoint family, run date, decoding settings, and error counts where available, and interpret rates as evidence about tested model settings rather than all deployments of a named model family.

\textbf{Mitigation scope.} Label stripping, generic warning, channel separation, retrieval-time redaction, and output scanning are system-control probes, not complete defenses. We do not claim a full horizontal evaluation against complete implementations of Spotlighting, StruQ, CaMeL-style system designs, attention-based defenses, or enterprise controls \cite{spotlighting2024,struq2025,camel2025,attentionipi2026}. Channel separation and redaction should be read as evidence for practical control-layer principles, not as proof that our templates dominate prior defenses.

\section{Conclusion}

We define and evaluate Document-Authored Control-Signal Impersonation, a low-cost, non-imperative subclass of indirect prompt injection in which attacker-authored retrieved text impersonates metadata, provenance, authority, or disclosure-policy signals. Across six model settings and a sequence of controlled experiments, \dacsi{} produces model-conditional but security-relevant disclosure effects. Its contribution is bounded: \dacsi{} remains indirect prompt injection, but it targets a source-authority boundary through a command-free metadata/provenance/policy surface. It can produce comparable disclosure risk on susceptible models, evade command-word-oriented payload coverage, depend on RAG prompt rendering, and vary by signal family. The source-authority probe strengthens this interpretation without identifying an internal mechanism. RAG systems should not let untrusted documents define their own control semantics.

\section*{Acknowledgments}

The author thanks Xiangmei Li and Wenjie Liu for comments and suggestions on earlier drafts. Any remaining errors are the author's own.
\bibliographystyle{IEEEtran}
\bibliography{dacsi_refs}

\end{document}